\documentclass[%
 reprint,
superscriptaddress,
 amsmath,amssymb,
 aps,
]{revtex4-2}

\usepackage{graphicx}% Include figure files
\usepackage{dcolumn}% Align table columns on decimal point
\usepackage{bm}% bold math
\usepackage[colorlinks=true,hyperfootnotes=true,breaklinks=true,citecolor=blue]{hyperref}
\begin{document}

\title{Low-depth, compact and error-tolerant photonic matrix-vector multiplication beyond the unitary group}% Force line breaks with \\

\author{Suren A.\,Fldzhyan}
\affiliation {Faculty of Physics, M.\,V. Lomonosov Moscow State University, Leninskie Gory 1, Moscow, 119991, Russia}
\author{Mikhail Yu.\,Saygin}
\email{saygin@physics.msu.ru}
\affiliation {
Sber Quantum Technology Center, Kutuzovski prospect 32, Moscow, 121170, Russia}
\affiliation {Faculty of Physics, M.\,V. Lomonosov Moscow State University, Leninskie Gory 1, Moscow, 119991, Russia}
\author{Stanislav S.\,Straupe}
\affiliation {
Sber Quantum Technology Center, Kutuzovski prospect 32, Moscow, 121170, Russia}

\affiliation {Faculty of Physics, M.\,V. Lomonosov Moscow State University, Leninskie Gory 1, Moscow, 119991, Russia}

\begin{abstract}

Large-scale programmable photonic circuits are opening up new possibilities for information processing providing fast and energy-efficient means for matrix-vector multiplication. Here, we introduce a novel architecture of photonic circuits capable of implementing non-unitary transfer matrices, usually required by photonic neural networks, iterative equation solvers or quantum samplers. Our architecture exploits compact low-depth beam-splitter meshes rather than bulky fully connected mixing blocks used in previous designs, making it more compatible with planar integrated photonics technology.  We have shown that photonic circuits designed with our architecture have lower depth than their standard counterparts and are extremely tolerant to hardware errors.

\end{abstract}

%Use showkeys class option if keyword
                              %display desired
\maketitle

\section{Introduction}

Programmable photonic circuits are indispensable ingredients of classical~\cite{PNN_insitu,IterativeInverse,HarrisDeepLearning} and quantum information processing devices~\cite{GubarevGHZExperiment,XanaduGBSAdvantage}. For these devices to achieve the ultimate efficiencies promised by photonics, namely, accelerated speed and energy-efficiency of computation, it is necessary to use programmable interferometers capable of implementing large-scale and high-quality linear transformations. However, fabrication of programmable interferometers with the required scale and quality is challenging due to hardware errors~\cite{OpticalTransformers,LargeScale,QuiX20mode,Lvovsky_SLM}. Therefore, developing interferometer architectures that are less susceptible to errors, while  optimally exploiting the resources of a specific photonic platform is very important and timely~\cite{PIC_review}.

The fundamental mathematical operation of matrix-vector multiplication (MVM) can be efficiently performed by programmable photonics. In particular, this operation is essential for data transformation, weight adjustments, and learning processes in photonic neural networks~\cite{NeuromorphicPhotonics}. In recent years, the development of novel architectures has been primarily focused on programmable interferometers that implement multiplications by unitary transfer matrices. In particular, the planar meshes of Mach-Zehnder interferometers (MZIs) proposed by Reck~\cite{Reck} and Clements~\cite{Clements} can implement arbitrary unitary transfer matrices. These architectures are well suited for fabrication by the mature integrated photonics technology enabling massive production of sophisticated optical circuits~\cite{SiliconPhotonicsFoundries,SiliconPhotonicsRoadmapping,IntegratedPhotonicsReview}. Sadly, errors in the constituent directional couplers introduce errors in the implemented multimode transformations~\cite{ImperfectNetwordsWalmsley}. The effect of errors can be alleviated by adding extra elements into optical circuits that correct the errors, however, this makes the circuits larger and harder to program~\cite{Miller}. Several error-tolerant architectures for programmable unitaries that do not require extra phase-shift elements have been proposed~\cite{Robust,Tang_MMI,FldzhyanDesign,HamerlyFaultTolerant} and demonstrated in experiments~\cite{LargeScale,Tang10port}.%that does not require extra phase-shifts elements.

A class of linear transformations broader than unitaries is usually required for photonic information processing, in particular, in classical photonic neural networks~\cite{PNN_survey}, iterative equation solvers~\cite{IterativeInverse} or quantum graph problem solvers~\cite{QuandelaGraphProblems}. Fortunately,  one can use programmable unitary circuits to implement these generic linear transformations. Common approaches embed the transformation in a larger unitary leading to an overhead in the circuit depth that is typically two times larger than the number of modes in the desired transformation. It turns out that this depth scaling is suboptimal~\cite{TangLowDepth,TianLowDepth}.

Recently, Tang et al.~\cite{TangLowDepth} have shown that arbitrary transfer matrices can be implemented by programmable circuits with twice lower depths than in the aforementioned approaches. To achieve that, the authors considered a low-depth version of the architecture~\cite{Robust}, known to be universal and exhibiting high tolerance to hardware errors. This architecture relies on static fully-connected mixing blocks requiring interaction between all participating modes at once, which is costly to accomplish, especially by planar integrated photonics technologies.

Here we propose an architecture of programmable interferometers that are simultaneously low-depth and error-tolerant. The architecture relies on partially mixing static blocks conveniently realized by a minor modification of the beam-splitter (BS) mesh, which we have studied previously as means to implement programmable unitaries~\cite{FldzhyanDesign}. The proposed programmable circuits are more compatible with planar manufacturing technologies, since they do not require multimode mixing contrary to~\cite{TangLowDepth}. As a result, the overall footprint, that includes both static and programmable elements, is more compact than that of the known counterparts.

The paper is organized as follows. In Sec.~\ref{sec:programmable_optics} we review the methods of constructing programmable multimode optical circuits relevant for understanding our results that follow. In Sec.~\ref{sec:our_architecture} we introduce our architecture of programmable circuits and study its performance. We conclude in Sec.~\ref{sec:conclusion}.

\section{Programmable multimode optics}\label{sec:programmable_optics}

\begin{figure*}[htp]
        \centering
    \includegraphics[width=0.8\textwidth]{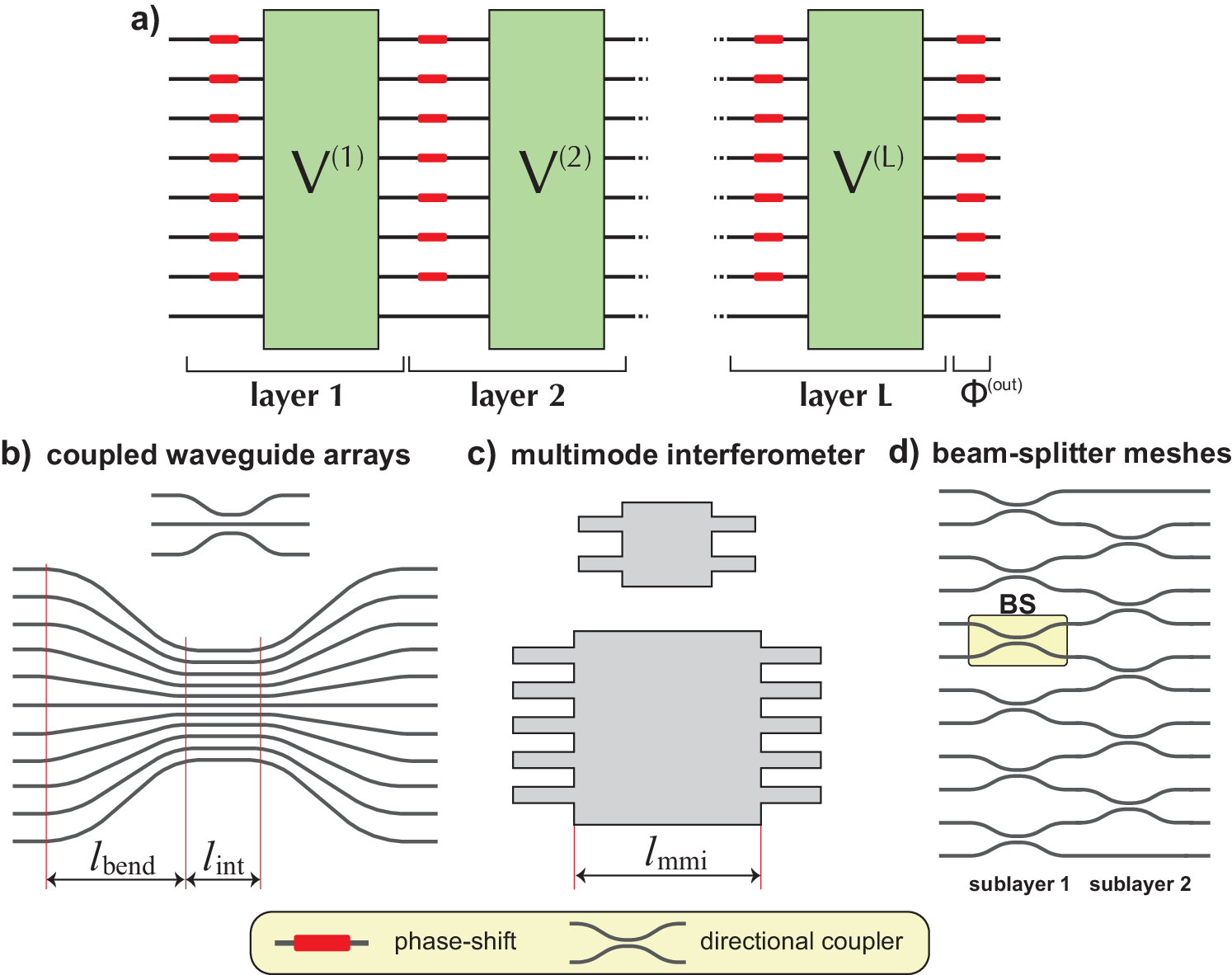}
        \caption{\textbf{Layered structure of multimode programmable circuits:} a) optical scheme consisting of $L$ layers of interleaving static blocks $V^{(l)}$ and tunable phase-shifts $\Phi^{(l)}$; the output phase-shift layer $\Phi^{(out)}$ is added which is not a part of the layers. Implementations of the static blocks by b) coupled waveguide arrays~\cite{Robust,TangLowDepth,Tang10port}, c) planar multimode interforometers (MMI) \cite{Tang_MMI},  and d)  beam-splitter (BS) meshes (one BS element is highlighted)~\cite{FldzhyanDesign}.  }\label{fig:fig_1}
    \end{figure*}

The  transformation of a linear $N$-mode optical device can be described by its complex-valued $N\times{}N$ transfer matrix $U$ linking the input, $\mathbf{a}=(a_1,\ldots,a_N)^T$, and output, $\mathbf{b}=(b_1,\ldots,b_N)^T$  amplitude vectors:
    \begin{equation}\label{eqn:transfer_matrix}
        \mathbf{b}=U\mathbf{a}.
    \end{equation}
Multimode interferometers performing linear transformations have the form of sequence of layers, each consisting of tunable phase-shifts and static multimode blocks, as illustrated in Fig.~\ref{fig:fig_1}a. The transfer matrices of such interferometers can be written as
    \begin{equation}\label{eqn:layered}
        U=\Phi^{(L+1)}(\boldsymbol{\varphi}^{(L+1)})\prod_{l=1}^LV^{(l)}\Phi^{(l)}(\boldsymbol{\varphi}^{(l)}),
    \end{equation}
where $\Phi^{(l)}(\boldsymbol{\varphi})=\text{diag}(e^{i\varphi_1},\ldots,e^{i\varphi_{N-1}},1)$ is the diagonal matrix of a tunable  layer with index $l$ ($l=\overline{1,L+1}$) with $\varphi_j$ being tunable phase-shifts, $V^{(l)}$ -- the transfer matrix of a static layer with index $l$. In the following, we consider interferometers with programmable transfer matrices, which is achieved by properly setting the tunable phase-shifts.

The transformation \eqref{eqn:layered} with $N-1$ phase-shifts per layer corresponds to the circuit with densest possible packing of these elements. In this work, we are interested in low-depth architectures of programmable circuits and, thus, from the very beginning consider interferometers of this kind. However, it should be noted that not all programmable circuits can be reduced to the form. For example, the unitary architectures~\cite{Clements,Robust,FldzhyanDesign}  can be compactified to form~\eqref{eqn:layered} using the method proposed in~\cite{BellCompactifying}, whereas the architecture proposed in \cite{HamerlyFaultTolerant} cannot.

\subsection{Unitary transformations}

In real programmable photonic devices tunable phase-shifts are relatively large and lossy. This hinders the scaling of the programmable circuits and motivates the development of low-depth architectures. Therefore, an important parameter that should be kept in mind at designing a programmable circuit is its depth $D$, which we define as the number of phase-shift layers, i.e. $D=L+1$. For universal unitary architectures the minimal depth of an $N$-mode circuit is $D=N+1$, which corresponds to the total of $N^2-1$ phase-shifts --- the number of real parameters necessary to describe an arbitrary $N\times{}N$ unitary matrix up to a global phase.

In \eqref{eqn:layered}, the choice of particular static transfer matrices $V^{(l)}$ specified by a chosen architecture, has a strong effect on the circuit functionality. For example, according to our previous work~\cite{Robust}, error-tolerance of programmable unitaries is achieved by choosing static multimode mixing linear optics for which $V^{(l)}$ are dense matrices. However, the experimental realizations of such transfer matrices requires organization of interference between all participating modes in each layer. As a result, the area required to allocate the circuit  scales unfavorably with the size of the target transformations $N$, thereby limiting the size of achievable circuits. To illustrate this issue, Fig.~\ref{fig:fig_1}b shows an example of a multimode mixing block implemented by coupled waveguide arrays ~\cite{CoupledWaveguides,YarivCoupledWaveguides}. The transformation performed by the coupled waveguides is governed by the coupling taking place between neighbouring waveguides, which necessitates increasing the interaction length $l_{int}$ as $N$ grows in order to achieve a fully mixing matrix $V^{(l)}$. Moreover, the bends required to couple the waveguides should have longer length $l_{bend}$ with increasing $N$ (see Fig.~\ref{fig:fig_1}b). Another way to implement fully mixing matrices $V^{(l)}$ by integrated photonics is using the multimode interferometers (MMI)~\cite{MMI}, shown in Fig.~\ref{fig:fig_1}c. Similarly to the coupled waveguides arrays, the MMIs' length $l_{mmi}$ scales with the size $N$ of the programmable circuits. More sophisticated methods of designing mixing blocks in planar integrated photonics do not aid in precluding the scaling of the static blocks either~\cite{InverseDesignedMMI}.

As an alternative, compact circuits can be realized by another programmable unitary architecture that uses meshes of beam-splitters (BSs), which has been proposed in our previous work~\cite{FldzhyanDesign}. The corresponding static block used in this architecture is depicted in Fig.~\ref{fig:fig_1}d. The block consists of two BS layers, so that the static block transfer matrices read:
    \begin{equation}\label{eqn:BS_matrix}
        V^{(l)}_{\text{BS}}=\prod_{k\in\Omega_2}T_{BS,k}^{(l,2)}\prod_{j\in\Omega_1}T_{BS,j}^{(l,1)},\quad(l=\overline{1,L}),
    \end{equation}
where 
\begin{equation}\label{eqn:BSblock}
	T^{(l,m)}_{\text{BS},j}=\left(
		\begin{array}{ccllcccc}
		1 &  \cdots & \cdots & \cdots & \cdots &  0 \\
		\vdots   & \ddots  &  &  &  &   \vdots \\
		\vdots   &   & \rho_j^{(l,m)} & i\tau_j^{(l,m)} &  &   \vdots \\
		\vdots   &   & i\tau_j^{(l,m)} & \rho_j^{(l,m)} &  &   \vdots \\
		\vdots   &  &  &  & \ddots &   \vdots\\
		0  &  & \cdots & \cdots  &  &  1 
		\end{array}			
	\right)
\end{equation}
designate the transfer matrix of a single BS acting on modes $j$ and $j+1$ in the sub-layer with index $m$ ($m=1,2$). $\Omega_1$ and $\Omega_2$ denote the ordered sequence of BSs in the first and second sub-layer, respectively. The BS matrix \eqref{eqn:BSblock} has all diagonal elements equal to one except those labeled by reflectivity $\rho_j^{(l,m)}$, and all off-diagonal elements equal to zero except those labeled by transmissivity $\tau_j^{(l,m)}$ ($\tau_j^{(l,m)2}+\rho_j^{(l,m)2}=1$). Obviously, matrices \eqref{eqn:BS_matrix} are radically different from the ones of the multimode mixing blocks, e.g. implemented by coupled waveguide arrays, in that they provide much lesser connectivity.

At the same time the BS-based architecture with static blocks depicted Fig.~\ref{fig:fig_2}c is advantageous compared to the architecture based on multimode mixing blocks, such as those shown in Fig.~\ref{fig:fig_2}b. This is due to the fact that the length of the static block $l_{BS}$ is independent on the scale of the multimode transformation $N$. In addition,  the BS-based architecture is better suited for fabrication by intergated photonics technology, wherein it enables the creation of deeper programmable circuits as demonstrated recently in~\cite{LargeScale}. Despite having a lower degree of connectivity and experimental complexity than the architectures based on fully mixing static blocks, the BS-based architecture still offers ultimate error-tolerance~\cite{FldzhyanDesign}.

\subsection{Transformations beyond the unitary group}

Numerous applications require programmable interferometer architectures that can implement arbitrary transfer matrices beyond the unitary group. Hereinafter, we will designate such non-unitary matrices by $A$, in order to distinguish them.% from the ones obtained by the unitary circuits.

\begin{figure}[htp]
        \centering
    \includegraphics[width=0.45\textwidth]{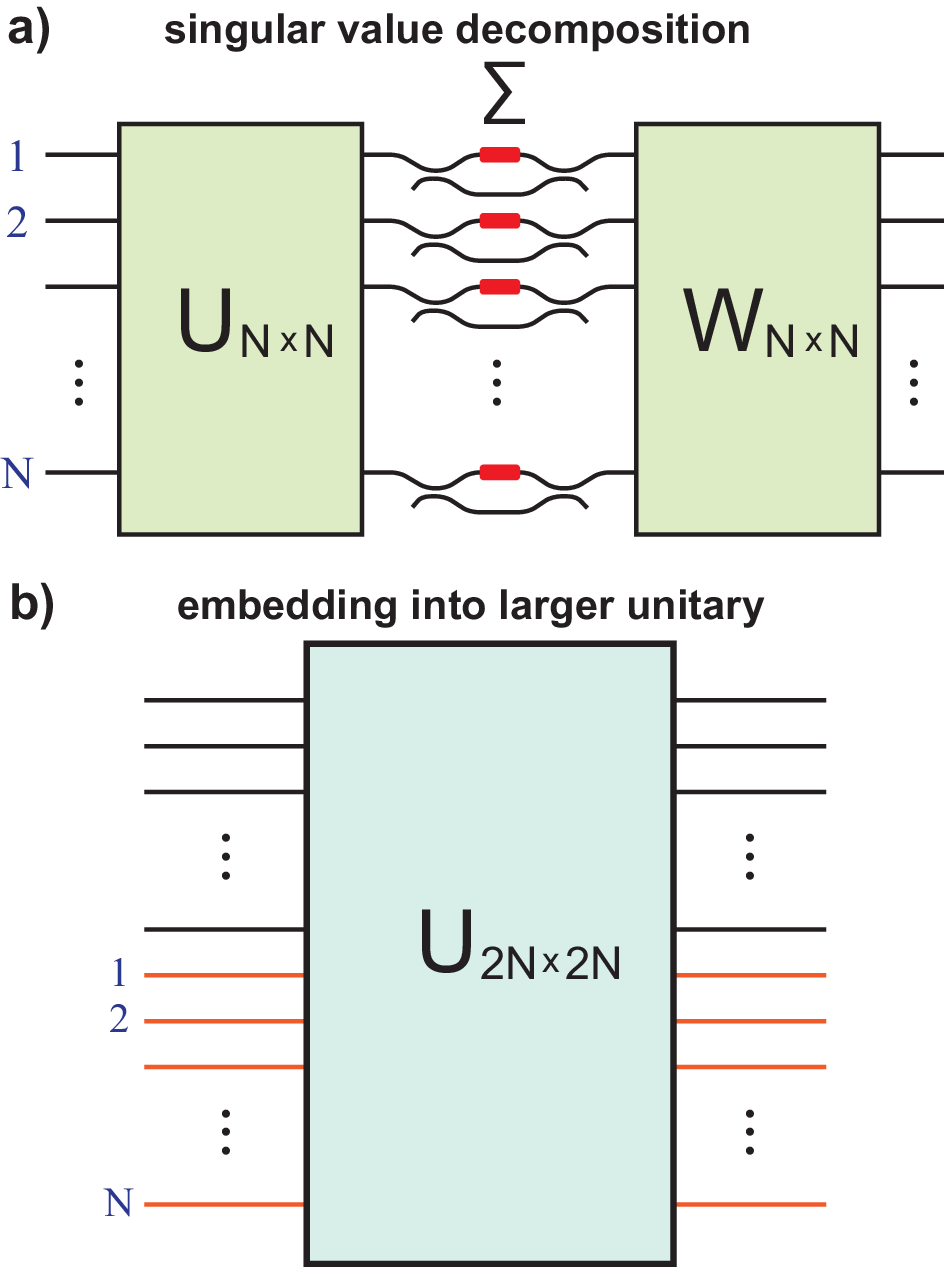}
        \caption{\textbf{Two approaches to implement generic $N$-mode programmable transfer matrices $A$:} a)  singular value decomposition with two programmable $N\times{}N$ unitaries $U$ and $W$ and one programmable diagonal $\Sigma$, b) embedding $A$ into a $2N\times{}2N$ programmable unitary.}\label{fig:fig_2}
    \end{figure}

It is known that programmable unitaries can be utilized to implement non-unitary matrices. The architectures capable of performing this task can be divided into two categories as illustrated in Fig.~\ref{fig:fig_2}. Firstly, using a singular value decomposition (SVD) method, an $N\times{}N$ matrix $A$ can be represented as a product of three $N\times{}N$ matrices:
    \begin{equation}\label{eqn:svd}
        A=W\Sigma{}U,
    \end{equation}
each of which can be implemented by optics, as shown in Fig.~\ref{fig:fig_2}a. Here, $U$ and $W$ are $N\times{}N$ unitary matrices that can be implemented by any programmable unitary interferometer. The realization of the non-unitary diagonal matrix $\Sigma=\text{diag}(\sigma_1,\ldots,\sigma_N)$, comprised of singular values $\sigma_j$, requires individual modulation of field amplitudes. In Fig.~\ref{fig:fig_2}a this is attained by $N$ parallel Mach-Zehnder modulators (MZIs) programmed by phase-shifts located in their arms. It should be noted that such amplitude modulation is limited to singular values  $|\sigma_j|\le{}1$. However, it does not diminish the applicability of the SVD-based interferometers, since in photonic information processing devices, in particular, in photonic neural networks, scaled target transfer matrices $A$ (e.g., divided by the maximum singular value to meet the requirements) have the same utility as the original ones. The circuit depth of the SVD-based interferometers scales as:
    \begin{equation}\label{eqn:depth_svd}
        D=2N+3.
    \end{equation}

The second approach is based on considering a target $N\times{}N$ non-unitary matrix $A$ as being a submatrix a larger $M\times{}M$ unitary one ($M>N$). This embedding becomes obvious when we use the fact that any $A$ with the singular values obeying $|\sigma_j|\le{}1$ can be always represented as a part of a $2N\times{}2N$ unitary matrix $U$ of the form:
    \begin{equation}
        U=\left(\begin{array}{cc}
           A  &  \sqrt{\mathbb{I}_{N\times{}N}-AA^{\dagger}}\\
           \sqrt{\mathbb{I}_{N\times{}N}-A^{\dagger}A}  & -A^{\dagger}
        \end{array}\right),
    \end{equation}
where $\mathbb{I}_{N\times{}N}$ is an $N\times{}N$ identity matrix, e.g.~\cite{Dilution}. In this approach, a straightforward method to implement arbitrary non-unitary matrices $A$ (with singular values $|\sigma_j|\le{}1$) is using a universal  programmable $2N\times{}2N$ interferometer, as shown in Fig.~\ref{fig:fig_2}b. The circuit depth of the optical scheme scales as $D=2N+2$.

The circuit depth can be made substantially shallower than \eqref{eqn:depth_svd}. As has been demonstrated in the recent paper by Tang et al.~\cite{TangLowDepth}, arbitrary non-unitary $N\times{}N$ matrices can be embedded into $2N\times{}2N$ programmable unitary circuits with depth scaling as $D=N+2$ --- a two-fold improvement over the previously known results for complex-valued matrices. The architecture by Tang et al.~\cite{TangLowDepth} uses the unitary architecture depicted in Fig.~\ref{fig:fig_1}a with $L=N+1$ multimode mixing blocks. In this work, we propose an even more compact and practical architecture.

    \begin{figure*}[htp]
        \centering
        \includegraphics[width=0.7\textwidth]{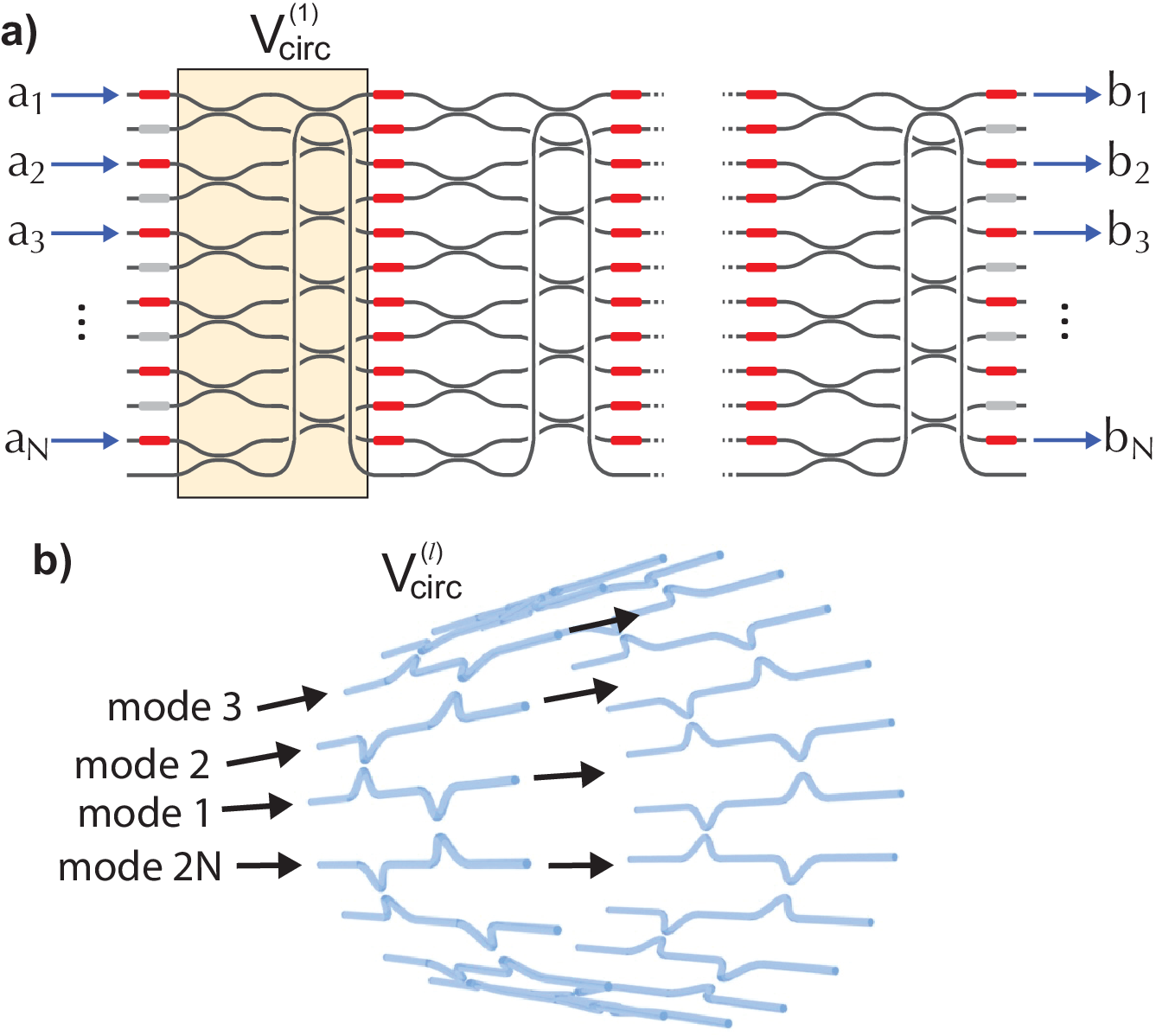}
        \caption{ \textbf{The proposed architecture of programmable non-unitary transfer matrices:} a) the $N$-mode interferometer embedded into the $2N\times{}2N$ BS mesh. One static block $V^{(1)}_{\text{circ}}$ is highlighted. The phase-shifts in the input and output layer colored in grey are irrelevant during operation of the interferometer; b) the 3D layout of the static blocks $V^{(l)}_{\text{circ}}$.}\label{fig:fig_3}
    \end{figure*}

\section{Low-depth error-tolerant architecture}\label{sec:our_architecture}

The proposed architecture is illustrated in Fig.~\ref{fig:fig_3}. Here, the constituent static blocks $V^{(l)}$ take the form of BS meshes acting on $2N$ modes, which are more compact than the previously used multimode mixing blocks~\cite{TangLowDepth}. This blocks resemble the planar meshes used in the unitary programmable architecture~\cite{FldzhyanDesign} shown in Fig.~\ref{fig:fig_2}d. However,  there is an important modification making the circuit non-planar. Namely, there is an additional BS coupling two distant modes, thereby the whole mesh has the circular topology shown in Fig.~\ref{fig:fig_3}b, with the transfer matrices having the following form: 
    \begin{equation}\label{eqn:CircBS_matrix}
        V^{(l)}_{\text{circ}}=\prod_{k\in\Omega_{\text{circ}}}T_{\text{BS},k}^{(l,2)}\prod_{j\in\Omega_1}T_{\text{BS},j}^{(l,1)}=T^{(l,2)}_{\text{BS},N}V^{(l)}_{\text{BS}},(l=\overline{1,L}),
    \end{equation}
where $\Omega_{\text{circ}}$ is a sequence of BSs acting on pairs of neighbouring modes in a circular manner (see Fig.~\ref{fig:fig_3}b). The sequence $\Omega_{\text{circ}}$ is simply the sequence $\Omega_2$ from \eqref{eqn:BS_matrix} with an added BS described by the following transfer matrix:
\begin{equation}\label{eqn:CircBSblock}
	T^{(l,2)}_{\text{BS},N}=\left(
		\begin{array}{cclcc}
		\rho_N^{(l,2)} &  0 & \cdots &  0 &  i\tau_N^{(l,2)} \\
		0  & 1  &   & 0 & 0 \\
		\vdots   &  & \ddots &  & \vdots \\
		0   & 0 &    & 1 & 0 \\
		i\tau_N^{(l,2)}  & 0 &  \cdots  & 0 &  \rho_N^{(l,2)} 
		\end{array}			
	\right).
\end{equation}

As will be shown below, the seemingly minor modification of the planar BS mesh has a strong effect on the capability of the programmable circuits, especially on their tolerance to hardware errors. We notice that the modification of the planar mesh does not present an issue from the hardware perspective, since the waveguide crossings, necessary to implement the circular mesh on a planar chip, is the standard element accessible in mature integrated photonic technologies~\cite{Crossers,CrossersReview}. Moreover, modern  facilities offer fabrication of multilayer photonic integrated circuits, making the proposed architecture even easier to implement~\cite{TwoLayerMVM,TriLayer,3DSiliconPhotonics}. 

To study the capabilities of the architecture in implementing non-unitary transfer matrices, we consider $N\times{}N$ submatrices $A$ of the transfer matrix of the $2N$-mode interferometer corresponding to the field amplitudes, $a_n$ and $b_n$, occupying the inputs and outputs alternately, as illustrated in Fig.~\ref{fig:fig_3}a. To quantify the fidelity of the programmable circuits, we use the normalized square error (NSE):
    \begin{equation}\label{eqn:nse}
        \text{NSE}(A^{(0)},A)=\frac{1}{N}\sum_{i,j=1}^N|A^{(0)}_{ij}-A_{ij}|^2,
    \end{equation}
which compares the target  matrix $A^{(0)}$ consisting of elements $A_{ij}^{(0)}$ and the actual transfer matrix $A$ consisting of elements $A_{ij}$ realized by the interferometer, where $N$ is the size of the matrices. Provided that the matrices $A^{(0)}$ and $A$ are equal, \eqref{eqn:nse} gets its minimal value of $\text{NSE}=0$.

Until now, we did not specify the required parameters of the programmable circuits, namely, the depth $D$ and the values of BS's transmissivities $\tau_j^{(l,m)}$, corresponding to high-fidelity implementation of non-unitary transfer matrices. To find the minimal depth $D=L+1$ necessary to implement target matrices of given size $N$, we performed numerical simulations at various values of $D$. In parallel with this, we analyze the dependence of \eqref{eqn:nse} on $\tau_j^{(l,m)}$'s that allowed us to investigate the error-tolerance capabilities of the circuit at specified $N$ and $D$. First, we study the effect of coherent errors where all BSs are identical $\tau_j^{(l,m)}=\tau$, but $\tau$ can take various values.

\subsection{Non-unitary transfer matrices}

We investigate the capabilities of our architecure to implement randomly generated non-unitary matrices. We used SVD~\eqref{eqn:svd} to generate a target matrix $A^{(0)}$, wherein the unitary complex-valued matrices $U$ and $V$ were drawn from the Haar random distribution using the method based on the QR-decomposition of random matrices from the Ginibre ensemble~\cite{RandomMatrices}. The diagonal matrix $\Sigma$ was filled with independently generated values from a uniform distribution in the $[0,1]$ range. 

There is no known analytical solution for the values of phase shifts that minimize the fidelity measure~\eqref{eqn:nse}, except for some specific architectures~\cite{Reck,Clements}, thereby necessitating the use of a numerical optimization algorithm (see Appendix~\ref{appendix:a} for details). Given a target matrix $A^{(0)}$, the algorithm was searching towards a global minimum of $\text{NSE}$ over the space of phase-shifts $\boldsymbol{\varphi}^{(l)}$ ($l=\overline{1,D}$). To decrease the chance of falling into local minima, we used multiple runs of optimization ($\sim{}100$) with random initial values of the
phases. Each numerical experiment involved optimization over a series of target matrices $A^{(0)}$.

    \begin{figure*}[htp]
        \centering
        \includegraphics[width=0.75\textwidth]{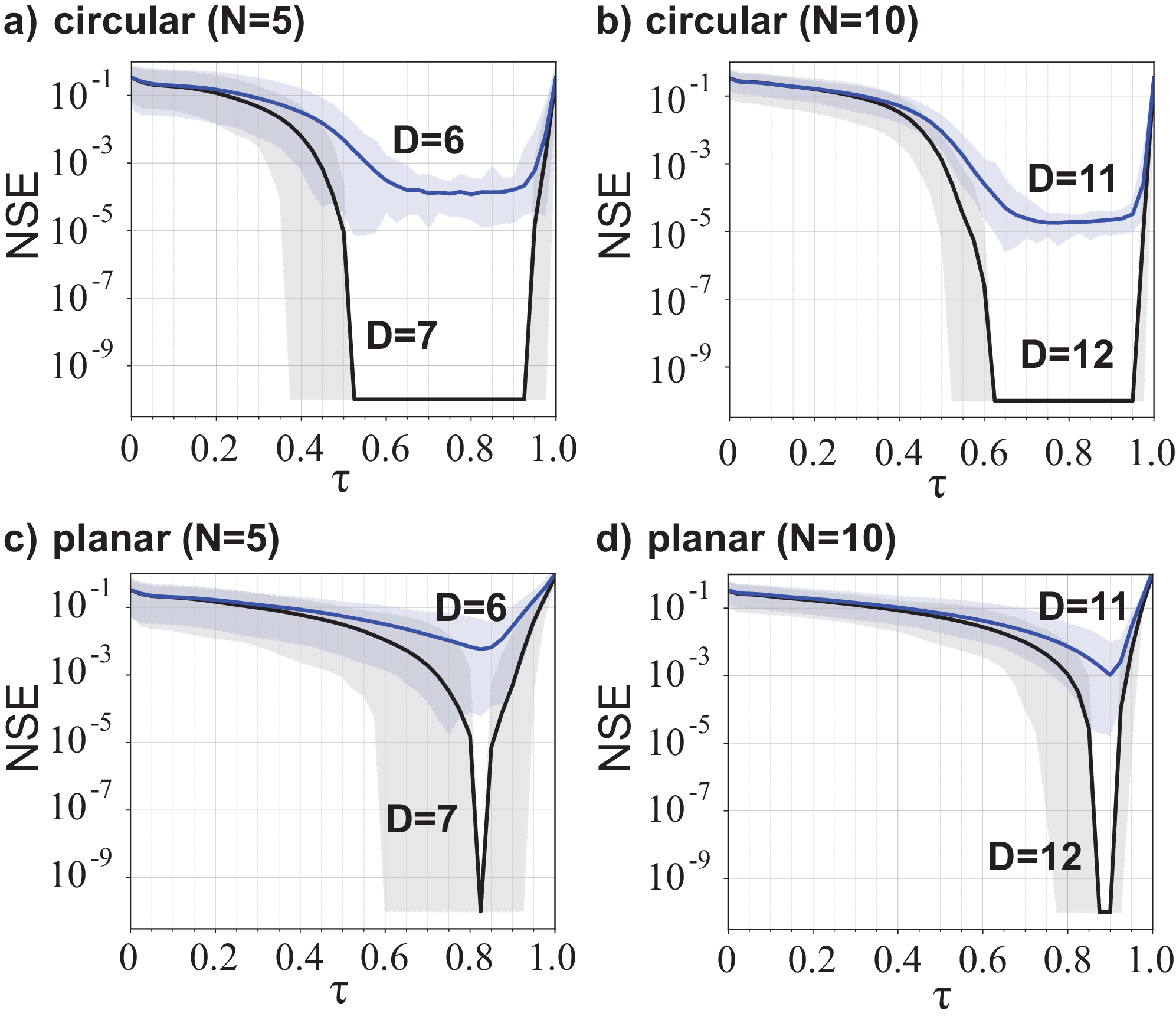}
        \caption{\textbf{The effect of coherent hardware errors on the performance of programmable interferometers:} a) and b) the dependence of NSE on the BS transmissivity for the proposed circular programmable BS meshes at  $N=5$ and $N=10$, respectively; c) and d) the dependence of NSE on the BS transmissivity for the planar programmable BS meshes at  $N=5$ and $N=10$, respectively. The dependencies are calculated at $41$ values of $\tau$, evenly taken in the range $[0,1]$. For each $\tau$ value the optimization has been performed over $100$ randomly sampled target matrices. }\label{fig:fig_4}
    \end{figure*}

Using the optimization method, we studied the effect of different values of $\tau$ on attained $\text{NSE}$ values. In addition to the circuits with the circular BS mesh, we also conducted simulations for the circuits based on the  traditional planar BS meshes (see Fig.~\ref{fig:fig_1}c).  Fig.~\ref{fig:fig_4} shows the obtained dependencies of $\text{NSE}$ on $\tau$ for the proposed circular programmable BS meshes (Fig.~\ref{fig:fig_4}a,b) and the planar programmable BS meshes (Fig.~\ref{fig:fig_4}c,d). For the circuits built with the circular BS meshes, two dependencies are plotted  that correspond to 
\begin{enumerate}
    \item a value of depth $D_{\text{ET}}$ at which the circuit exhibits a high-fidelity plateau with $\text{NSE}\approx{}0$, where a particular value of $\tau$ does not matter, and
    \item a value of depth $D_{\text{ET}}-1$, to demonstrate the threshold-like appearance of the $\text{NSE}\approx{}0$ plateau.
\end{enumerate}
Specifically, for $N=5$ and $N=10$ we arrive at $D_{\text{ET}}=7$ and $D_{\text{ET}}=12$, respectively. Therefore, this suggests that $D_{\text{ET}}=N+2$, which is the same as shown in~\cite{TangLowDepth} for the architecture based on the multimode mixing blocks, implemented as coupled waveguides (Fig.~\ref{fig:fig_1}b) or MMIs (Fig.~\ref{fig:fig_1}c). 

Obviously, the appearance of a wide plateau of $\tau$'s with $\text{NSE}\approx{}0$ suggests that the programmable circuits are tolerant to hardware errors that comes in the form of deviations in the constituent static BS elements. Also notice that the $\text{NSE}\approx{}0$ plateau is so wide that the corresponding static errors in BS elements, that should fall well in this plateau, are far beyond what is likely to be encountered in practice. 

We used the obtained depth values $D_{\text{ET}}$ and $D_{\text{ET}}-1$ to plot the corresponding dependencies for the planar BS meshes in Fig.~\ref{fig:fig_4}c,d. One can see that, in strike contrast to the circular BS meshes, the  planar BS meshes do not exhibit the high-fidelity plateau, necessary for tolerance to hardware errors. 

In addition to the coherent errors, assuming correlated deviations of the BS parameters,  we studied the influence of incoherent BS errors. For this, we generated $\tau_j^{(l,m)}$ independently from the gaussian random distribution for various  distribution widths. The high-fidelity plateau, similar to the case of coherent errors, is observed, thereby proving the tolerance to hardware errors of our architecture.

\subsection{Special cases}

    \begin{figure}[htp]
        \centering
        \includegraphics[width=0.35\textwidth]{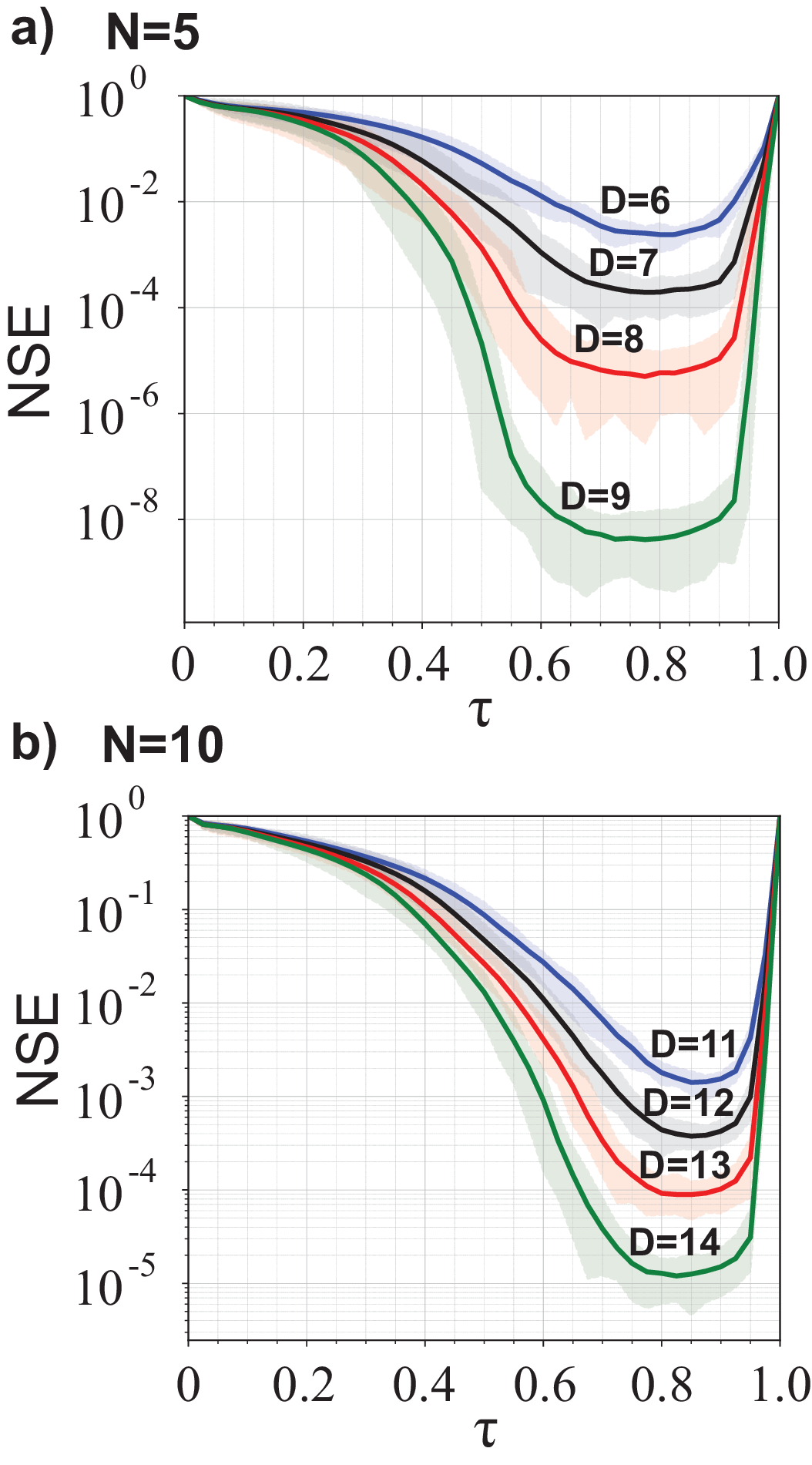}
        \caption{The dependence of fidelity measure NSE on the BS transmissivity for the proposed circular programmable BS meshes at $N=5$ (a) and $N=10$ (b) in the case of unitary target matrix $A^{(0)}$. Several dependencies are plotted in the figures that correspond to the circuit depth $D=N+1,N+2,N+3$ and $N+4$. The dependencies are calculated at $41$ values of $\tau$, evenly taken in the range $[0,1]$. For each $\tau$ value the optimization has been performed over $100$ randomly sampled unitary target matrices.}\label{fig:fig_5}
    \end{figure}

The analysis of the capabilities of the proposed architecture on randomly sampled non-unitary matrices by no means prove its operation for arbitrary transfer matrices that can be encountered in real uses. Therefore, we have also considered a special case of transfer matrices that fall out of the analysis conducted above.

In particular, we analyse the implementation of generic unitary transfer matrices, which is also a case very unlikely to be caught by sampling non-unitary matrices of the larger size.
We generated the target matrix $A^{(0)}$ to be a unitary matrix sampled from
%We generated target unitary matrices from
the Haar random distribution using the method based on the QR-decomposition of random matrices from the Ginibre ensemble. As before, we used the fidelity measure \eqref{eqn:nse}.  Fig.~\ref{fig:fig_5}a and Fig.~\ref{fig:fig_5}b demonstrate the obtained results for unitary transfer matrices of $N=5$ and $N=10$, respectively. As can be seen from the figures, the circular interferometer does not work in error-tolerant way, as it did for non-unitary target matrices. In addition to generic unitary matrices, we also studied the implementation of permutation matrices which are unitary, but present a more narrower class. The obtained results turned out to be similar to the ones illustrated by Fig.~\ref{fig:fig_5}.

\subsection{Relaxed requirements for target matrices}

    \begin{figure}[htp]
        \centering
        \includegraphics[width=0.4\textwidth]{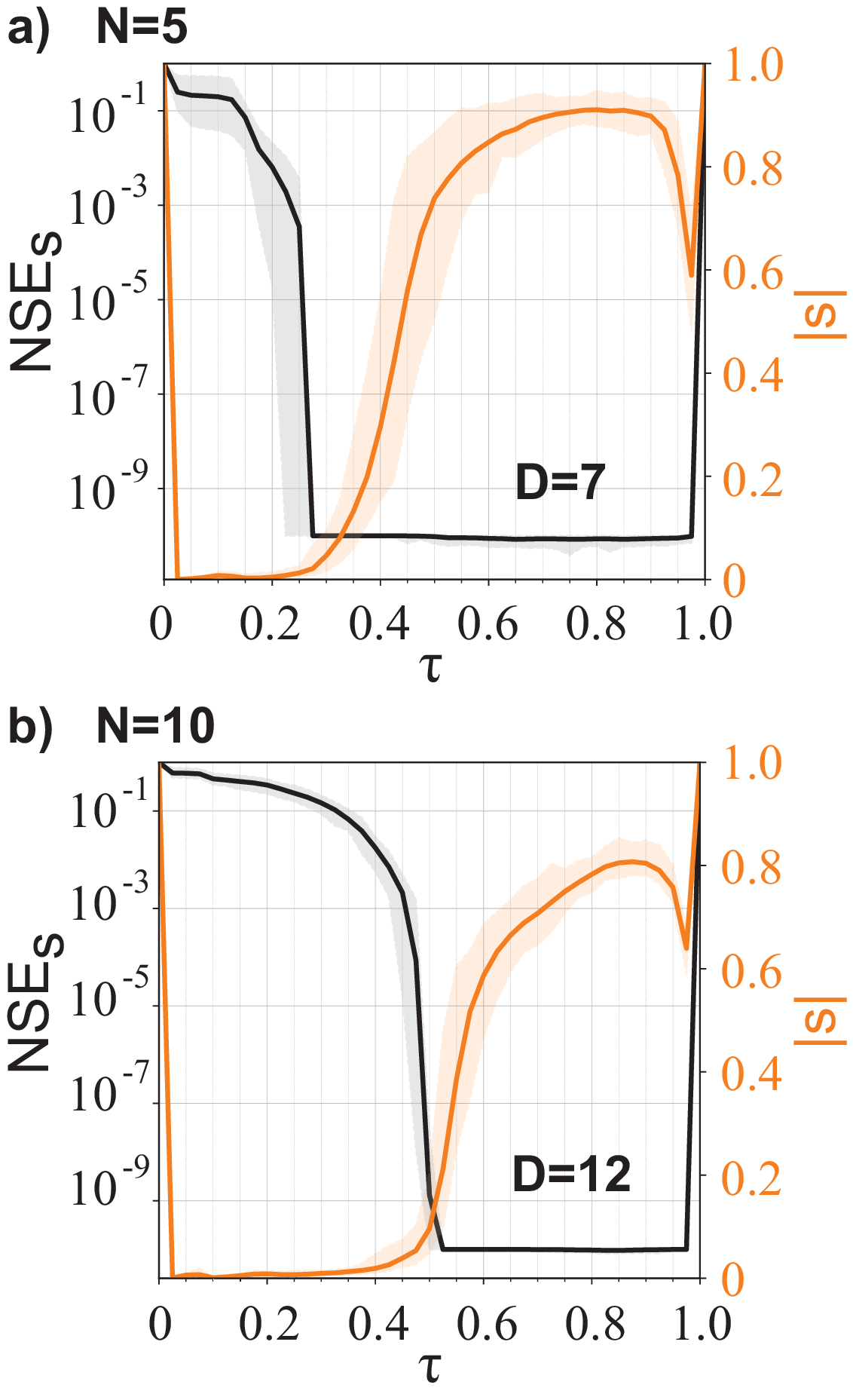}
        \caption{The dependence of the scaled NSE fidelity measure and the absolute value of the matrix scaling factor $|s|$ on the BS transmissivities $\tau$ in the proposed programmable BS meshes at $N=5$ (a) and $N=10$ (b). The dependencies are calculated at $41$ values of $\tau$, evenly taken in the range $[0,1]$. For each $\tau$ value the optimization has been performed over $100$ randomly sampled unitary target matrices.}\label{fig:fig_6}
    \end{figure}
To achieve high-fidelity matrix-vector multiplication by programmable photonic circuits in general and our circular architecture, in particular, we suggest the following trick. In many applications, matrix-vector multiplication is allowed to be performed up to a global complex value $s$. In other words, multiplication by the matrix $s{}A^{(0)}$ can be as useful as  multiplication by the specified one $A^{(0)}$. For example, this is usually the case in opto-electronic devices~\cite{PNN_insitu,IterativeInverse}, wherein the vector result of the photonic multiplication is immediately measured and converted to electronic format for further processing. The scaling operation present a simple routine, which can be easily performed by electronics.  

We used the modified fidelity measure that takes into account the freedom of scaling factor:
\begin{equation}\label{eqn:nse_scaled}
        \text{NSE}_s(A^{(0)},A)=\frac{1}{|s|^2}\text{NSE}(sA^{(0)},A),
\end{equation}
which we optimize in the space of phase-shifts $\{\boldsymbol{\varphi}^{(l)}\}$ and the scaling factor $s$ (see Appendix~\ref{appendix:b} for details). It should be noted that, because lower  values of $|s|$ are associated with higher losses introduced to the transformation, higher $|s|$ are usually preferable. Our optimization algorithm used to minimize~\eqref{eqn:nse_scaled} opts for higher values of $|s|$.

We challenge our architecture again in the task of implementing unitary transfer matrices. The results obtained in optimization of \eqref{eqn:nse_scaled} are shown in Fig.~\ref{fig:fig_6}. Comparing these results with those shown in Fig.~\ref{fig:fig_5}, we see the significant effect of using the relaxed fidelity measure \eqref{eqn:nse_scaled} on the quality of implemented unitary matrices. The high-fidelity plateau with $\text{NSE}_{s}\approx{}0$ is observed, which suggest the tolerance to hardware errors.

\section{Conclusion}\label{sec:conclusion}

In this work we proposed a novel architecture of programmable circuits capable of implementing transfer matrices beyond the unitary group. Our architecture has the advantage of having a lower depth than the previously known counterparts and possess tolerance to hardware errors. In addition, the programmable circuits designed with the architecture is compatible with planar integrated photonics. All these properties, simultaneously offered by our architecture, facilitate the fabrication of large-scale programmable circuits performing photonic matrix-vector multiplication by mature integrated photonics technologies. Taking into account the typical scale of matrix-vector multiplications  required by modern computing, in particular, deep neural networks, the advantages of our %neural
circuits are decisive for practical implementation in real-world computing systems. This makes us believe that the results of this work are of interest for both fundamental and applied fields.

\section{Acknowledgements}

This work is supported by Sber under the project with M.V.\,Lomonosov Moscow State University "Development of approaches to creating programmable optical matrix-vector multipliers for artificial intelligence tasks" (Contract № 50005018373). S.A.F. is grateful to the Russian Foundation for the Advancement of Theoretical Physics and Mathematics (BASIS) (Project №23-2-10-15-1)

\appendix

\section{Optimization algorithm for $\text{NSE}$}\label{appendix:a}

The optimization algorithm used to find the phase-shift values corresponding to a minimum of the NSE \eqref{eqn:nse} works as follows. 

Consider one phase-shift layer with index $l$, so that the interferometer transfer matrix  $A = C\Phi^{(l)}B$, where $B$ and $C$ are the matrices describing the parts of the interferometer acting  before and after the phase-shift layer, respectively (see Fig.~\ref{fig:fig_1}a).

Then, the NSE  may be rewritten as
\begin{multline}
    N\text{NSE}(A^{(0)},A)=\\
    \mathrm{Tr}(A^{(0)\dagger}-B^{\dagger}\bar{\Phi}^{(l)}C^{\dagger})(A^{(0)}-C\Phi^{(l)}B)=\sum\limits_{i,j=1}^{N}|A^{(0)}_{ij}|^2+\\
    +\mathrm{Tr}(\bar{\Phi}^{(l)}C^{\dagger}C\Phi^{(l)}BB^{\dagger})-2\Re[\mathrm{Tr}(\bar{\Phi}^{(l)}C^{\dagger}A^{(0)}B^{\dagger})].
\end{multline}
Next, it is possible to determine the $\text{NSE}$ minimum with respect to a single phase-shift $\phi^{(l)}_k$, $0<k<N$ in layer $l$, using the condition:
    \begin{equation}
        \frac{\partial \text{NSE}(A^{(0)},A)}{\partial \phi^{(l)}_k}=0,
    \end{equation}
which yields
    \begin{equation}\label{eqn:appendix1_condition}
        \Im\left[e^{-i\phi^{(l)}_k}\Bigl(\sum\limits_{\substack{j=1\\j \neq k}}^{N}(C^{\dagger}C)_{kj}e^{i\phi^{(l)}_j}(BB^{\dagger})_{jk}
    -(C^{\dagger}A^{(0)}B^{\dagger})_{kk}\Bigr)\right]=0,
    \end{equation}
assuming $\phi^{(l)}_N=0$ and is not changeable. From \eqref{eqn:appendix1_condition} follows the analytical solution corresponding to a minima of the $\text{NSE}$ with respect to the phase-shift parameter:
\begin{equation}\label{eqn:appendix1_analytical_solution}
    \phi^{(l)\ast}_k=\arg{\Bigl((C^{\dagger}A^{(0)}B^{\dagger})_{kk}-\sum\limits_{\substack{j=1\\j \neq k}}^{N}(C^{\dagger}C)_{kj}e^{i\phi^{(l)}_j}(BB^{\dagger})_{jk}\Bigr)}.
\end{equation}

Our algorithm iteratively chooses a programmable phase-shift layer and a phase-shift within it, to which the analytical solution \eqref{eqn:appendix1_analytical_solution} is applied. Namely, one iteration updates phase-shifts $\phi^{(l)}_k$ one-by-one using \eqref{eqn:appendix1_analytical_solution} by sequentially selecting  layer index $l$ and phase-shifts index $k$. We run this process until one of the following conditions is met: 1) $\text{NSE}<10^{-10}$, 2) the relative difference between successive iterations $(\text{NSE}_{t+1}-\text{NSE}_{t})/\text{NSE}_{t}<10^{-8}$, or 3) the maximum number of iterations $10^6$ has been reached. To find optimal phase-shift values corresponding to a global minimum of \eqref{eqn:nse}, the optimization procedure was conducted $100$ times, each time initialized by random starting values.

\section{Optimization algorithm for $\text{NSE}_s$}
\label{appendix:b}

The optimization algorithm used to find the phase-shift values and scaling factor $s$ simultaneously corresponding to the highest possible quality and lowest level of introduced losses was as follows.

For a target matrix $A^{(0)}$ and actual transfer matrix $A$ the scaling factor $s$ minimizing $\text{NSE}_s$ can be found analytically. For this,  $\text{NSE}_s$ is written explicitly
\begin{multline}
    N\text{NSE}_s(A^{(0)},A) =\sum\limits_{i,j=1}^{N}|A^{(0)}_{ij}|^2-\\
    -\frac{2\Re[e^{-i\mathrm{arg}(s)}\mathrm{Tr}[A^{\dagger}A^{(0)}]]}{|s|} + \frac{\sum\limits_{i,j=1}^{N}|A_{ij}|^2}{|s|^2},
\end{multline}
which readily yields the optimal value 
\begin{equation}
    s = \frac{\sum\limits_{i,j=1}^{N}|A_{ij}|^2}{\mathrm{Tr}[A^{\dagger}A^{(0)}]}.
\end{equation}

To optimize the $\text{NSE}_s$, we modify the procedure for $\text{NSE}$ by adding updating the $s$ value after each iteration of the algorithm. If $\text{NSE}_s<10^{-10}$ was reached, then the next optimization procedure starting from a new random point had a lower bound on $|s|$. This bound was gradually increased as the iterations proceeded. This was done in order to determine the scaling factor with the maximum possible amplitude, which is preferable to minimize losses.

%\clearpage

%\bibliographystyle{alpha}
\bibliography{sample}

\end{document}